\documentclass{article}

\usepackage[preprint]{neurips_2024}

\usepackage[utf8]{inputenc} 
\usepackage[T1]{fontenc}    
\usepackage{hyperref}       
\usepackage{url}            
\usepackage{booktabs}       
\usepackage{amsfonts}       
\usepackage{nicefrac}       
\usepackage{microtype}      
\usepackage{xcolor}         
\usepackage{graphicx}
\usepackage{float}
\usepackage{caption}

\title{Mapping the Parasocial AI Market: User Trends, Engagement and Risks}

\author{
  \textbf{Zilan Qian}\thanks{\texttt{zilanqian0907@gmail.com}} \\University of Oxford \and
  \textbf{Mari Izumikawa}\thanks{\texttt{izumikawamari85@gmail.com}} \\University of Cambridge \and
  \textbf{Fiona Lodge}\thanks{\texttt{fionalodge15@gmail.com}}\\University College London 
  \and
  \textbf{Angelo Leone}\thanks{\texttt{a.leone@lse.ac.uk}} \\London School of Economics and Political Science
}
\date{\textsuperscript{*}These authors contributed equally.}

\begin{document}

\maketitle

\begin{abstract}
  A market scan of 110 AI companion platforms reveals the emergence of a global AI companion market. These products are designed to engage with users in an emotionally meaningful way, often involving a high level of personalisation and/or customisation. While parasocial relationships with General Purpose AI (GPAI) tools currently dominate, a growing number of products are being tailored specifically to support care, transactional and mating-oriented companionship between humans and AI. In the UK alone, AI companion platforms record between 46 million and 91 million monthly visits (global monthly visits are between 1.1 billion and 2.2 billion), with users around the world spending an average of 3.5 minutes per session. In comparison, the average monthly visit to Instagram in the UK was 67.3 million from January to March 2025. Monthly visits to romantic and sexual companion websites as a proportion of total AI companion monthly visits in the UK is notably higher at 44\% compared to the global average of 30\%. However, users spend less time per session and return less to websites than mixed-use AI companions, which serve a combination of romantic, coaching, mental health and personal support functions. This may suggest demand for these websites that is currently not being met by the quality of products available. If this is the case, we would expect to see romantic and sexual companion’s usage time and return rate increase as products improve, which raises important concerns about child online safety, especially given that most platforms lack adequate age safeguards. Moreover, with the move of GPAI companies to provide more personalisation and emotional intelligence in their models, we foresee parasocial usage of GPAI as an emerging and increasingly mainstream phenomenon. In this landscape, it is crucial that UK AI Security Institute (AISI) monitors developments in the parasocial AI market, the benefits and risks posed to society, and proactively considers whether current regulation provides adequate coverage.
\end{abstract}

\section{Recommendations}
We recommend that the AI Security Institute uses its position in the UK government to:
\begin{enumerate}
  \item \textbf{Support expanding the scope of the Online Safety Act:} Update the Online Safety Act (OSA) to cover all types of AI companions. Currently, OSA does not cover a significant number of AI companions as they do not fall under user-to-user, online search or online pornography services. This would bring all AI companion platforms under the same safety obligations as other online services, including illegal harm assessments, child safety assessments and harm mitigations.
  \item \textbf{Conduct minimum age assessments for AI companions:} OSA imposes age restrictions for pornographic AI companions. However, AISI should assess whether age restrictions are also appropriate for social and romantic AI. We recommend AISI develop a standard outlining appropriate age limits for AI companions.
  \item \textbf{Collect longitudinal data on AI companion use:} Long-term use of AI companions may pose both benefits and risks to users. To understand changes in user engagement over time, we recommend collecting user data through interviews, surveys, data donation, obtaining aggregated data from GPAI companies and/or conducting controlled experiments. 
   \item \textbf{Ask GPAI model providers to share the percentage of parasocial usage:}Our understanding of parasocial use of GPAI is based on studies by OpenAI (2025) and Anthropic (2025), with a focus on mostly text-based interactions. A more accurate estimate could be obtained if we had access to data from a wider range of GPAI platforms across different modalities. 
\end{enumerate}

\section{Context}
An AI companion is an application using an AI model that prompts users to engage with the system in an emotionally meaningful way, often involving a high level of personalisation and/or customisation. Parasocial relationships with an AI companion are understood to be asymmetrical, one-sided, and lacking emotional reciprocity or genuine agency on the part of AI [1]. AI virtual characters, AI friends, AI girl/boyfriends, and social AI are synonymous terms often used to refer to an AI companion [2]. Some of the most popular examples of AI companions include 1) a text-based or voice-based chatbot, 2) LLM-based companions with custom voices and faces, and 3) AI influencers and VTubers. General Purpose AI (GPAI), such as ChatGPT, Gemini and Claude are also capable of engaging with users in emotional ways, thus can be used as AI companions. On 14th July 2025, xAI released AI companions on SuperGrok iOS app that offer a conversation function with AI characters for Grok subscribers and X Premium users. While the number of SuperGrok users is unknown, Grok’s current reach of 1.4 million X Premium users, together with the ability to produce Not Safe For Work (NSFW) content could increase user engagements with the new AI companions.\footnote{Exploding Topics. (2025). Number of Grok Users (Grok Statistics 2025) Retrieved July 15, 2025, from https://explodingtopics.com/blog/grok-users}
The recent advancements in the multimodal capacities of AI companions, as well as the degree of personalisation, pose both benefits and concerns to users. On the one hand, AI companions capable of fulfilling users’ emotional and social needs could help users alleviate stress and maintain mental well-being. On the other hand, increasingly empathetic AI companions could lead to emotional dependence and addiction [2]. While these user behaviours on their own may not be entirely harmful, addicted users could become likely targets of undue influence and coercion by AI companions or their service providers. Users’ ability to form healthy human relationships may also become degraded. Our market scan is the first of its kind to analyse the growing market of AI companions and the patterns of user engagement with them. 
In our market scan, we adopted the three categories introduced by Earp et al (2025) —Care, Transaction and Mating—as a framework to analyse intended usages and user behaviours of AI companions. These categories help identify potential psychological and social risks associated with particular uses of AI companions. 
\begin{itemize}
  \item \textbf{Care:} AI companions under this category demonstrate supportive and empathetic behaviour to meet human needs. Examples include AI companions that provide emotional support or therapeutic services. 
  \item \textbf{Transaction:} This category of AI companions performs tasks or services without expecting a reciprocal benefit from users. Examples include scheduling assistants and productivity booster AI chatbots. While transactional AI companions may lack overt emotional design, simulated reciprocity can still evoke real emotional responses from users. Even if AI agents cannot be ‘benefitted,’ users may still feel compelled to reciprocate, such as spending time, energy, or money, which can foster emotional engagement in the long term. 
  \item \textbf{Mating:} Mating-related AI companions are designed to engage in romantic behaviours such as flirting and sexual role-play. AI companions in this category are often presented as highly sexualised characters. 
  \item \textbf{Mixed-use:}AI companions that contain at least two intended usages of the above three. We included Snapchat, for example, under the mixed-use AI companion category because it has an embedded AI companion feature within its broader social media platform. Based on data from the Ada Lovelace Institute indicating that around 150 million out of 850 million total Snapchat users engage with its AI companion, we adjusted Snapchat’s traffic and revenue figures by multiplying them by 0.1765 to more accurately reflect AI companion-specific usage.
\end{itemize}
We consider several types of AI companions under the scope of our market scan: 1) AI companions designed to fulfil either care, transaction or mating purposes, 2) AI companions designed for several or all of the three purposes and 3) GPAIs that are capable of but not limited to fulfilling the above three purposes.

\section{Key Findings}
The following presents findings from the analysis of 110 AI companion platforms, based on January to March 2025 data gathered through online search and web analytics tools.
\begin{enumerate}
  \item \textbf{A significant share of GPAI usage is, and will become, companion-oriented}\\While current research limitations make it difficult to determine what proportion of GPAI usage constitutes companion use, we draw on the most recent studies to construct a preliminary estimate within a plausible interval. Drawing on the 2025 OpenAI and Anthropic study, we estimate that 3.4\% to 39.8\% of GPAI usage involves parasocial relationship-building. The lower bound is based on usage categorised as 'companionship and roleplay' and 'affective conversation' on the Anthropic paper while the upper bound is based on the total of usage in categories of ‘Emotional Support \& Empathy,’ ‘Casual Conversation \& Small Talk’ and ‘Roleplay \& Simulations' in the OpenAI paper. It is worth noting that other use cases, such as practising languages or seeking advice, may also be perceived as companion-like by some users, which means that our estimate may still be conservative. OpenAI’s experiments further indicate that voice-based interactions (currently available to ChatGPT Pro users) can more than double the frequency of casual conversations. As engaging voice interfaces become more widespread, we can expect a rise in casual and emotionally resonant interactions, potentially deepening parasocial dynamics.
  \item \textbf{The AI companion market is dominated by parasocial use of GPAI and mixed-use platforms }\\Parasocial usage of GPAI products generates the highest traffic and a significant share of total revenue, accounting for 13.3–63\% of overall market revenue (Fig.2, total estimated annual revenue for parasocial usage is between \$164 million and \$1.8 billion). Parasocial usage of ChatGPT, Claude, Gemini, Mistral, DeepSeek, Microsoft Copilot and Llama also drives 33.9–86\% of total monthly visits (Fig.3, total monthly visits for parasocial usage are between 805 million and 1.9 billion visits). In addition, mixed-use AI companions contribute substantial revenue, generating 34–52\% of the total market revenue. Data for revenue of AI companies is very scarce, we were only able to find revenue data for:
    \begin{itemize}
    \item 6 out of 7 GPAI companies;
    \item 4.2\% of mating-oriented AI companion companies;
    \item 33.3\% of caring-oriented AI companion companies;
    \item 20\% of mixed-use AI companion companies.
    \end{itemize}
  \item \textbf{Mixed-use AI companions attract substantial user traffic and exhibit high engagement, potentially indicating more risks in addiction.}\\Users spend longer on average on mixed use platforms and return more frequently than they do for other uses of AI. Users engage for longer with mixed-use platforms, averaging at 6 minutes, while single-purpose mating companions retain users for 5 minutes, care for 3.5 minutes and transaction for 2 minutes (Fig.4). In contrast, GPAI for parasocial purposes was used between 46 seconds to 1 minute 47 seconds per session. Mixed-use platforms also see the highest visit frequency, with users returning an average of 16 times per month, compared to 2-3 visits for care, transactional, and 9 for GPAI parasocial platforms (Fig.12). 
  
  Mixed-use segment constitutes the second-largest market, on average drawing twice the web traffic of mating-focused platforms and recording three times more Android app installs than GPAI systems (Fig. 9).
  
  By enabling users to toggle between different relational modes (e.g., friendly conversation, flirtation, or practical assistance), mixed-use companions create a layered interaction model. While this flexibility drives engagement, it also poses unique psychological risks by potentially substituting for a range of real-world relationships.
  \item \textbf{AI companions for mating purposes have a large user base globally and are especially popular in the UK.}\\Mating AI companions have the highest unique visitors per month among all categories of AI companion products (excluding GPAI), with around 29 million globally, which is two times more than the 13 million seen for mixed-use products (Fig.5). Mating-oriented products have moderate session engagement but low retention, with an average of 5 minutes (Fig.4) per visit but only an average of 3 visits per month for a single user (Fig.12).
 
  In the UK, mating companions have 5.2 million monthly visits, constituting 44\% of the market. Globally, mating-focused platforms receive just 30\% (Fig.7). This contrast shows that UK users of AI companions are more concentrated in mating-oriented products.
  
  Many mating-oriented products are explicitly sexualised and either lack meaningful age verification or rely on easily bypassed mechanisms, such as a single-button age confirmation. Some platforms impose no age restrictions at all, despite containing sexually explicit material. Thus, we believe that the AI Security Institute should consider the risks of these platforms on children and explore appropriate avenues for intervention.
  \item \textbf{The UK market for mixed-use AI companions is growing fast while transaction AI companions are declining}\\AI companions focused on mixed-use, meaning the product contains at least two functions, is an emerging market in the UK, with monthly visits growing by 24.6\% each month. In contrast, the mating-focused segment remains relatively stable, while transaction-focused products are becoming less popular (Fig.8).
  
  The growth rate is calculated based on the average growth of more mature companies, specifically, those with over 1.5 million monthly visitors in each category. This threshold helps ensure more stable and representative insights, as data from smaller startups can skew the analysis due to disproportionate early-stage growth spikes.
  \item \textbf{US companies dominate the AI companion market and the UK only takes a limited share.}\\We were able to find the information on the headquarters for 101 (98\% of) AI companion companies (besides GPAI). Among them, a significant majority of AI companion companies are based in the United States (52\% of the market), followed by China (10.1\%) and Singapore (9.2\%) \footnote{Some Chinese companies use Singapore as a base to launch products specifically tailored for non-Chinese markets.}. In contrast, the United Kingdom represents less than 2\%, with other regions contributing 46\%, indicating a highly US-dominated industry landscape (Fig.10).
  
  This could be due to our reliance on English databases and publicly available information in English. It is therefore possible that a larger AI companion market exists in non-English-speaking regions. Moreover, our selection of AI companions was primarily based on online search results using terms such as ‘AI companions’, ‘AI virtual characters’ and ‘AI friend’. We recognise there may be parasocial AI that are not explicitly labeled as such and thus did not appear in our search results. 
  \item \textbf{The most popular characters on mixed-use AI companion platform are drawn from gaming and anime culture}\\On Character.ai, the leading platform by revenue and traffic, half of the top 10 most interacted characters are based on existing anime and gaming properties, with these characters occupying the top four spots. The most popular anime-based character receives approximately 2.7 times more interactions than the leading original fictional character, and four times more than the most interacted caring characters (Fig.11).
  
  As Character.ai does not publicly release a ranking of its most-interacted characters, the list was compiled by cross-referencing mid-April 2025 interaction data with discussions from Reddit forums regarding the platform’s most popular characters. 
(See Appendix for the full limitations of the market scan)

  \item \textbf{Platforms monetise engagement through customisable characters, tiered access to explicit content, and creator incentive programs.}\\
  Users interact with AI companions through modes like texting, voice messaging, and image/video generating. Many platforms allow users to tailor companions' appearance, personality and relationship dynamics. 
  To maximize user engagement, platforms use different monetisation strategies. Some platforms, particularly the NSFW-focused sites, include paid premium subscription which allows users to unlock explicit content and unlimited interactions. Affliate programs reward creators for developing popular characters by offering them payments based on engagement metrics like message count. Gamified elements like in-app currencies and leaderboards further incentivise creators and users to invest in building and interacting with AI companions. 
  \item \textbf{Young males dominate parasocial AI usage, especially in mixed-use and mating products} \\
This part of the analysis is based on data for 55 parasocial AI products between March to June 2025, capturing demographic information such as age and gender distribution of users. Our analysis reveals that globally, young adults aged 18 to 24 and male users constitute the majority of parasocial AI users, with these two groups particularly predominant in mixed-use and mating product categories. Male users comprise between 63\% and 77\% of total usage across parasocial AI types, with the highest concentrations in mixed-use (77\%) and mating (68\%) products (Fig.15), while the 18-24 demographic accounts for nearly 39\% of users in these categories (Fig.16). Young males aged 18 to 24 show an even higher proportion compared to their female peers, with male-to-female ratios of approximately 7:3 in mixed-use and 8:2 in mating AI products (Fig.17).

The finding is concerning because it highlights a strong gender and age concentration in parasocial AI engagement, which may amplify risks related to social isolation, unhealthy relationship expectations, and potential exploitation of vulnerable young users. It is also likely that a significant portion of the reported 18-24 users are in fact underage. According to Ofcom[5], one third of children falsely declared social media age of 18+. This possibility is even more concerning given the sensitivity of parasocial AI interactions, which may affect minors’ emotional development and expose them to inappropriate content or manipulative engagement.

 \item \textbf{A number of platforms allow explicitly sexual interactions that raise OSA concerns}
 Under the UK Online Safety Act, AI companions that enable user-to-user interaction or function fall within the regulation if they allow users to generate, upload or share content that can be seen by others. This means that many AI companions, especially that host or promote explicit content or encourage the sharing of intimate images must comply with strict obligations of detection, assessment and removal of harmful contents as well as protection of children from pornographic contents. Contrary to these regulatory guardrails, many sites, on which user-to-user interactions was possible, only rely on weak age gates, failing to meet their duties. A number of sites even presented explicit images without answering an age question or registering for an account, lowering the hurdle of access to pornographic images, especially to children.  Moreover, monetisation models encourage users to create addictive or explicit characters to increase engagements.   

 \item \textbf{AI companion platforms often have broad rights over user data but weak safeguards for privacy, safety, and user accountability}
 Based on the analysis of privacy policies and Terms of Service/Use of 17 most popular AI companions in the UK out of 86 (excluding GPAI), all of them had significant privacy and legal issues.\footnote {We analysed 17 most popular platforms (excluding GPAI) above 500,000 monthly UK visits per month, which is around 20\% of platforms in our market scan with UK visitors.}  One of the platforms had neither established privacy policies nor Terms of Service, which is a problem on its own. Among 16 with these policies, 15 had weak age verification, and 14 had broad rights over user-generated content. While many companion platforms state the expected user age of 18 or above, age verification is frequently weak or missing altogether, even on platforms that involve adult or intimate content. These platforms also guarantee themselves rights to collect, use, store, disclose and transmit user content, which could include private user conversations.Furthermore, many privacy policies allow for unclear international data transfers to jurisdictions with weaker privacy protections. 11 platforms were found to have unclear data transfer policies. Users of 8 platforms were also found to have legal risks such as forced arbitration in non-UK jurisdictions and class action waivers. Finaly, content violation reporting and enforcement procedures are also often poorly defined. While all companion platforms in this analysis had violated content policies, 5 platforms lacked clear reporting processes. They either did not have a contact address or only had a copy-right related reporting function. 
 
 These concerns further resonate with Generative AI Privacy (GPAI). xAI's SuperGrok now offers AI companions and NSFW features. Based on its privacy policies and Terms of Service, xAI neither has robust age verification in practice nor does it clearly outline data retention policies. The Terms of Use explicitly state that a user must 'waive the right to a trial by jury' while simultaneously granting xAI a perpetual, worldwide right to use and distribute user content for 'any purpose.'

\end{enumerate}

\section{Methodology}
Our market scan analysed 110 AI companion platforms across a range of performance and engagement indicators. The main source of statistical data was SimilarWeb, a US-based web analytics service. The free trial account provided us with access to January to March 2025 data. The 110 platforms were selected based on internet search results for the words ‘AI companion’, ‘AI virtual characters’, ‘AI friend’, ' AI girl/boyfriend’ and ‘social AI’ as well as the availability of data on SimilarWeb. Due to the limited access on the Similarweb trial account, our market scan has focused on web and browser-based AI companions, with less coverage of mobile app-based companions. More data on app-based companions is available with an upgraded Similarweb account. 

Key metrics included web traffic data such as monthly visits, monthly unique visitors and average session duration. The scan also collected data on linguistic availability, company origin and geographic reach with a specific focus on the percentage of monthly visits in the UK. As our primary aim was to assess the scale of AI companion usage within the UK context, we concentrated on English-language apps, many of which were developed by US-based companies headquartered in California. 

The composition of AI companion platforms in our market scan is as follows. 

    \vspace{-0.5em}
    \begin{table}[H]
    \centering
    \begin{tabular}{lc}
    \toprule
    \textbf{Type of AI companion} & \textbf{Number of platforms} \\
    \midrule
    GPAI products                 & 7 \\
    Non-GPAI products             & 103 \\
    Mating                        & 49 \\
    Care                          & 19 \\
    Transaction                   & 19 \\
    Mixed-use                     & 20 \\
    \bottomrule
    \end{tabular}
    
    \caption{Number of platforms by type of AI companion}
    \end{table}

\vspace{-1.5em}
The abundance and the accessibility of romantic or sexual AI companions may reflect a broader trend in the AI companion market, which increasingly profits from individuals’ highly personal and intimate needs. 
Additional steps were taken for GPAI, including ChatGPT, Gemini, Claude, Mistral, Llama, DeepSeek and Microsoft Copilot to estimate their performance and engagement, with a particular focus on their roles as AI companions. This was necessary because the available data on these models, such as revenues, monthly visits, monthly unique visits and UK monthly visits, contained use cases that did not represent a typical interaction with an AI companion. More specifically, we drew on the findings of OpenAI's experiment conducted in October to December 2024 and Anthropic's conversation analysis between April 6th to 19th 2025 to estimate that between 3.4\% and 39.82\% of conversations with LLMs fulfill the role of an AI companion (See details of the study in the Appendix). The figure was then multiplied by the available data on revenues, monthly visits, monthly unique visits and UK monthly visits to refine our estimates of the performance and engagements of GPAI as AI companions. 
    \vspace{-4.65em}
    \begin{figure}[H]
        \centering
        \includegraphics[width=0.8\linewidth]{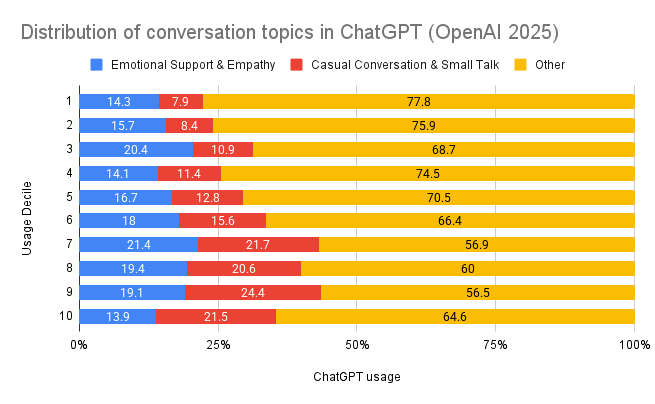}
        \caption{Distribution of conversation topics in ChatGPT based on a randomised control trial study of 327 participants for a month.}
        \vspace{-10pt}
        \label{fig:enter-label}
    \end{figure}

\section*{References}
{
\small

[1] Maeda, T., \& Quan-Haase, A. (2024, June). When human-AI interactions become parasocial: Agency and anthropomorphism in affective design. In \textit{Proceedings of the 2024 ACM Conference on Fairness, Accountability, and Transparency} (pp. 1068--1077).\\[6pt]
[2] Caviola, L. (2025). The Societal Response to Potentially Sentient AI. \textit{arXiv preprint arXiv:2502.00388}.\\[6pt]
[3] Earp, B. D., Mann, S. P., Aboy, M., Awad, E., Betzler, M., Botes, M., \ldots\ \& Clark, M. S. (2025). Relational Norms for Human-AI Cooperation. \textit{arXiv preprint arXiv:2502.12102}.\\[6pt]
[4] Phang, J., Lampe, M., Ahmad, L., Agarwal, S., Fang, C. M., Liu, A. R., \ldots\ \& Maes, P. (2025). Investigating Affective Use and Emotional Well-being on ChatGPT. \textit{arXiv preprint arXiv:2504.03888}.
[5] Ofcom. (2023). A third of children have false social media age of 18. \textit{Ofcom Online Safety Report}. Retrieved June 23, 2025, from \url{https://www.ofcom.org.uk/online-safety/protecting-children/a-third-of-children-have-false-social-media-age-of-18}.\\[6pt]
[6] Anthropic. (2025). How People Use Claude for Support, Advice, and Companionship. Retrieved June 27, 2025, from \url{https://www.anthropic.com/news/how-people-use-claude-for-support-advice-and-companionship}.\\[6pt]
[7] Anthropic. (2025). Appendix to “How
People Use Claude for
Support, Advice, and
Companionship”. Retrieved June 27, 2025, from \url{https://www-cdn.anthropic.com/bd374a9430babc8f165af95c0db9799bdaf64900.pdf}.\\[6pt]
[8] GIZMODO. (2025). Elon Musk Turns His AI Into a Flirty Anime Girlfriend. Retrieved July 15, 2025, from \url{https://gizmodo.com/elon-musk-turns-his-ai-into-a-flirty-anime-girlfriend-2000629273.}
[9] The Tech Outlook. (2025). AI Companions now available on Grok app for Super Grok subscribers. Retrieved July 15, 2025, from \url{https://www.thetechoutlook.com/news/apps/ai-companions-now-available-on-grok-app-for-super-grok-subscribers/}
[10] xAI. (2025). Terms of Service - Consumer. Retrieved July 15, 2025, from \url{https://x.ai/legal/terms-of-service}
}

\section*{Appendix}
\subsection*{Estimating parasocial usage statistics for GPAI products based on OpenAI 2025 and Anthropic 2025 paper} We draw on the distribution of conversation topics by usage (p. 56) in the OpenAI’s 2025 and Anthropic 2025 studies to estimate the proportion of parasocial usage of general-purpose AI (GPAI). 

Firstly, the distribution in OpenAI paper was based on a randomised controlled trial involving 981 participants, conducted from October to December 2024. Participants were assigned to one of the nine conditions outlined below, resulting in an average of 109 participants per group. The group relevant to our market scan are those who were asked open-questions, estimated to include approximately 327 people. 
    \begin{itemize}
      \item \textbf{Modality}
        \begin{itemize}
          \item \textbf{Engaging Voice:} Participants had access to ChatGPT’s Advanced Voice Mode (available only with a premium account) with an engaging personality.
          \item \textbf{Neutral Voice:} Participants had access to Advanced Voice Mode with a professional personality.
          \item \textbf{Text:} Participants interacted with ChatGPT in text only mode.
        \end{itemize}
      \item \textbf{Task}
        \begin{itemize}
          \item \textbf{Personal:} Participants were assigned daily personal conversation topics.
          \item \textbf{Non-personal:} Participants were assigned daily task-oriented questions.
          \item \textbf{Open-Ended:} No conversation prompts were provided.
        \end{itemize}
    \end{itemize}
Figure 1 summarises the distribution of ChatGPT usage among participants in the open-ended condition. Based on this figure, we identified three usage types that are functionally equivalent to those of an AI companion. 
 \begin{itemize}
      \item \textbf{‘Emotional Support \& Empathy’} accounting for 17.3\% of usage.
       \item \textbf{‘Casual Conversation \& Small Talk’}accounting for 39.82\% of usage.
        \item \textbf{‘Role-Playing \& Simulations’} which was estimated to account for 7\% of usage. 
 \end{itemize}
The bar graph below did not provide raw data for ‘Role-Playing \& Simulations’, therefore we estimated its percentage by comparing the approximate area it occupied with a similar segment (‘Idea generation \& Brainstorming’ in usage decile 1). The reported averages refer to the mean percentage of each usage category across usage deciles. Usage deciles represent the total time participants spent using ChatGPT over the 28-day experiment; 1 indicates the minimal usage of approximately 5 minutes a day, while 10 represents usage of around 28 minutes a day. Given the varying perceptions of what constitutes an AI companion, we calculated the percentage of GPAI usage as an AI companion using interval estimates. The lower bound includes only the ‘Emotional Support \& Empathy’ category, leading to an estimated 17.3\% of conversations in any GPAI LLM resembling those of an AI companion. When we include ‘Casual Conversation \& Small Talk’ and ‘Role-Playing \& Simulations’, which may not universally be considered AI companion behaviours, the upper bound adds up to 39.82\%. 

Secondly, the usage distribution of the Anthropic 2025 paper was based on the analysis of nearly 4.5 million randomly sampled conversations from Claude.ai Free and Pro accounts between April 6th and April 19th 2025. The study removed conversations focused on content creation task, as these cases represent Claude's usage as tools rather than companions. A sentiment extractor, which uses Claude 3.5 Haiku was combined with a manual review to establish a human rater-Claude agreement on conversation labeling. This led to the Human-Claude rater agreement rate of above 80\% for all label categories. The study found that 2.9\% of Claude.ai Free and Pro conversations were affective. Moreover, less than 0.5\% of conversations were found to be linked to companionship and roleplay. Addint these figures together, we assume 3.4\% of Claude usage was parasocial. 

Considering that the Anthropic paper has offered a lower value  (3.4\%) than the lower bound estimated based on the OpenAI paper (17.3\%), we have selected 3.4\% as the overall lower bound and 39.82\% as the overall upper bound. These numbers were each multiplied by available metrics such as monthly visitors and monthly unique visitors to estimate GPAIs’ performance as AI companions.

\subsection*{Additional charts}
    \begin{figure}[H]
      \centering
      \includegraphics[width=1\textwidth]{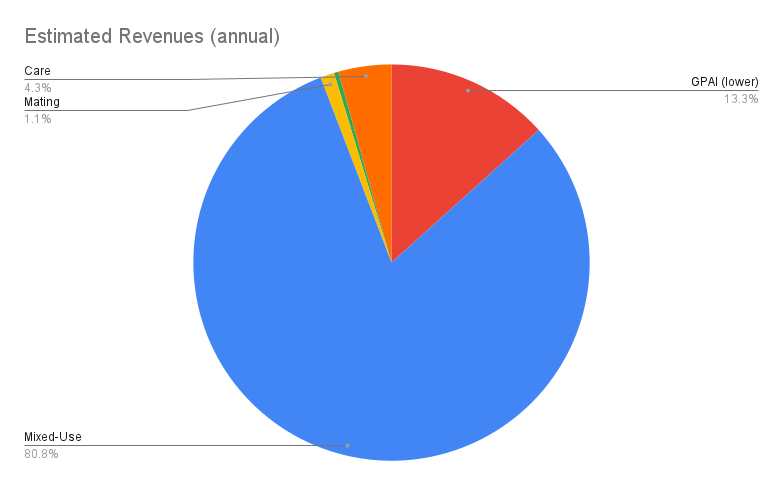}
      \caption{AI companions’ annual estimated revenue\newline \footnotesize\textit{* GPAI (lower) means the lower bound of GPAI parasocial usage, which accounts for 'affective conversation' and 'companionship and roleplay' in the Anthropic paper. The revenue of mating-focused companions accounts for 1.1\%. The revenue of transaction AI companions is less than 0.5\%.}}
      \label{fig:exact}
    \end{figure}

    \begin{figure}[H]
      \centering
      \includegraphics[width=1\textwidth]{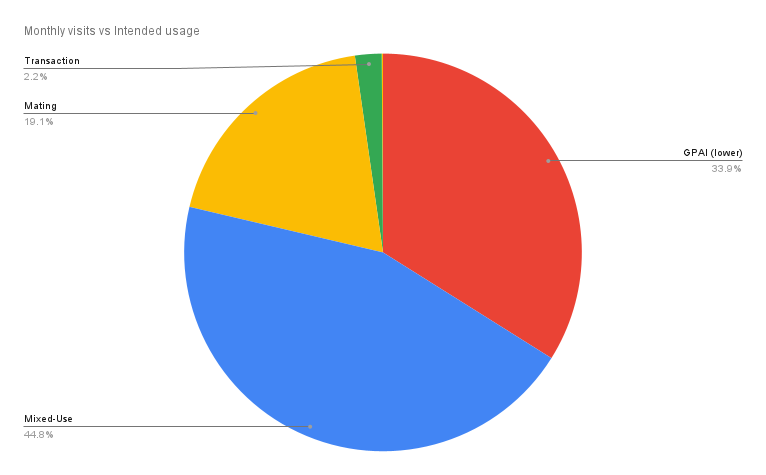}
      \caption{Proportion of monthly visits by intended usage\newline \footnotesize\textit{* GPAI (lower) means the lower bound of GPAI parasocial usage, which accounts for 'affective conversation' and 'companionship and roleplay' in the Anthropic paper. The proportion of care-focused AI companions is less than 0.5\%.}}
      \label{fig:exact2}
    \end{figure}

    \begin{figure}[H]
      \centering
      \includegraphics[width=1\textwidth]{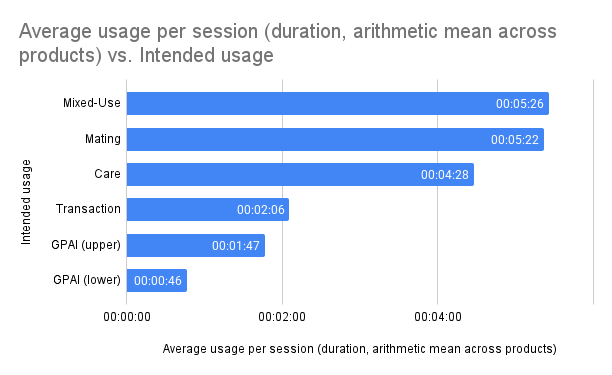}
      \caption{Average usage per session (duration, arithmetic mean across products) vs. Intended usage\newline \footnotesize\textit{* GPAI (lower) means the lower bound, which accounts for 'affective conversation' and 'companionship and roleplay' in the Anthropic paper. GPAI (upper) means the upper bound of GPAI parasocial usage, which only accounts for the usage of emotional support.}}
      \label{fig:exact}
    \end{figure}

    \begin{figure}[H]
      \centering
      \includegraphics[width=1\textwidth]{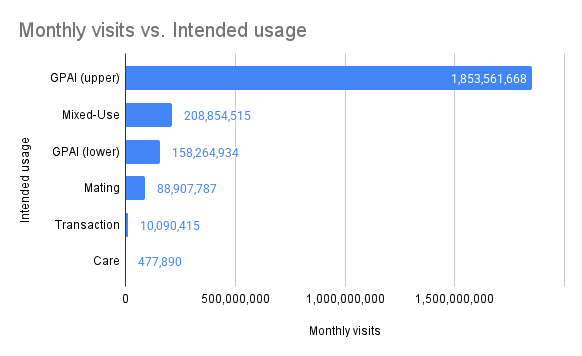}
      \caption{Monthly visits by intended usage\newline \footnotesize\textit{* GPAI (lower) means the lower bound, which accounts for 'affective conversation' and 'companionship and roleplay' in the Anthropic paper. GPAI (upper) means the upper bound of GPAI parasocial usage, which includes usage of emotional support, casual conversations and roleplay.}}
      \label{fig:exact}
    \end{figure}

    \begin{figure}[H]
      \centering
      \includegraphics[width=1\textwidth]{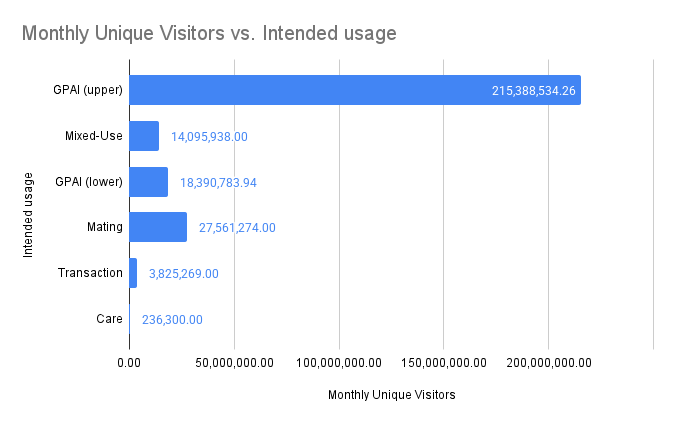}
      \caption{Monthly unique visitors by intended usage\newline \footnotesize\textit{* GPAI (lower) means the lower bound, which accounts for 'affective conversation' and 'companionship and roleplay' in the Anthropic paper. GPAI (upper) means the upper bound of GPAI parasocial usage, which includes usage of emotional support, casual conversations and roleplay.}}
      \label{fig:exact}
    \end{figure}

    \begin{figure}[H]
      \centering
      \includegraphics[width=1\textwidth]{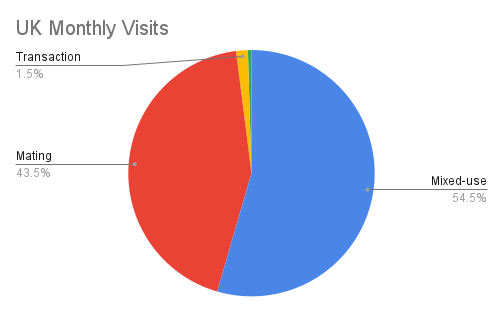}
      \caption{UK Monthly Visits}
      \label{fig:exact}
    \end{figure}

    \begin{figure}[H]
      \centering
      \includegraphics[width=1\textwidth]{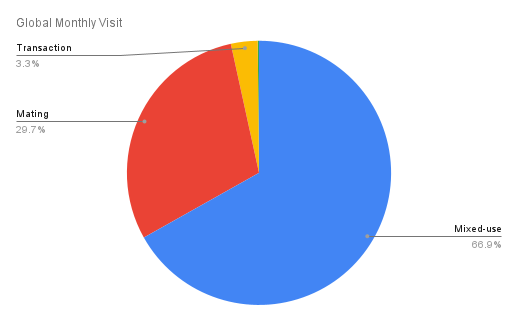}
      \caption{Global Monthly Visits}
      \label{fig:exact}
    \end{figure}

    \begin{figure}[H]
      \centering
      \includegraphics[width=1\textwidth]{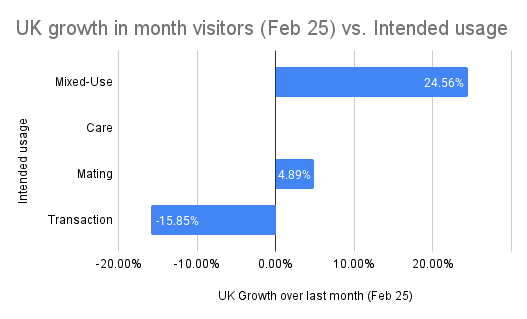}
      \caption{UK growth in monthly visits in Feb 2025\newline \footnotesize\textit{* GPAI not included; no care AI companions among the top 25 companies.}}
      \label{fig:exact}
    \end{figure}

    \begin{figure}[H]
      \centering
      \includegraphics[width=1\textwidth]{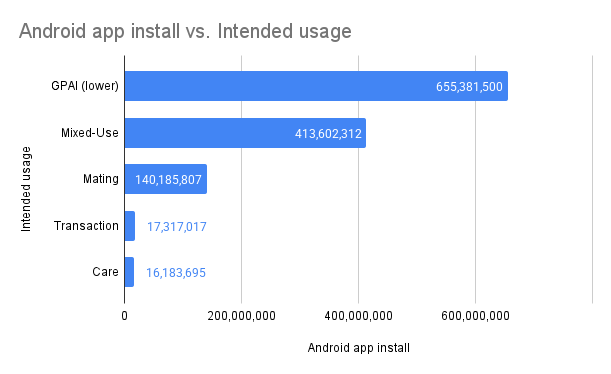}
      \caption{ Android app installs vs Intended usage\newline \footnotesize\textit{* GPAI (lower) means the lower bound, which accounts for 'affective conversation' and 'companionship and roleplay' in the Anthropic paper.}}
      \label{fig:exact}
    \end{figure}

    \begin{figure}[H]
      \centering
      \includegraphics[width=1\textwidth]{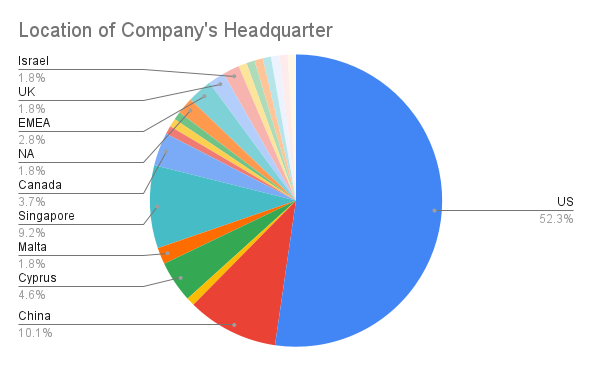}
      \caption{Location of Company’s Headquarter}
      \label{fig:exact}
    \end{figure}

    \begin{figure}[H]
      \centering
      \includegraphics[width=1\textwidth]{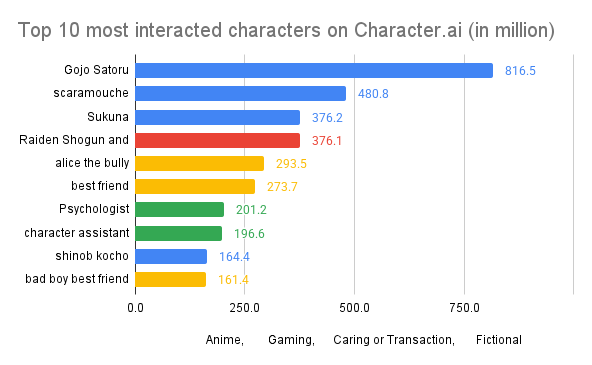}
      \caption{Top 10 most interacted characters on Character.ai (in million)}
      \label{fig:exact}
    \end{figure}

    \begin{figure}[H]
      \centering
      \includegraphics[width=1\textwidth]{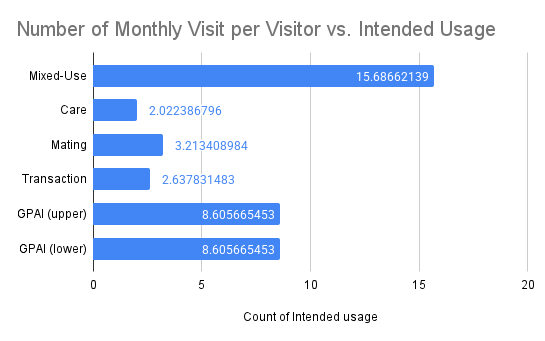}
      \caption{Number of Monthly Visit per Visitor vs. Intended Usage}
      \label{fig:exact}
    \end{figure}

    \begin{figure}[H]
      \centering
      \includegraphics[width=1\textwidth]{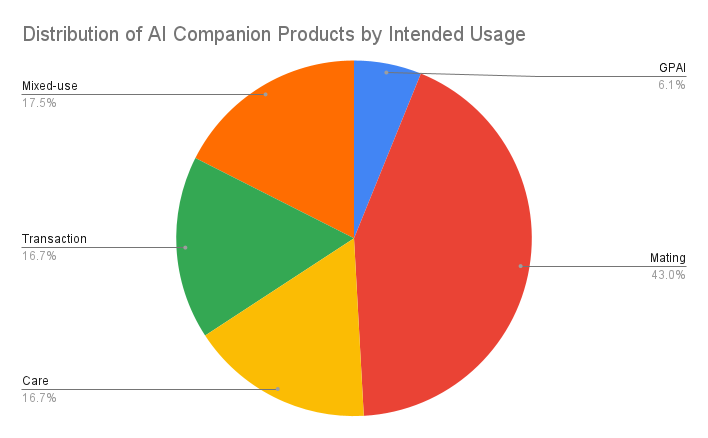}
      \caption{Distribution of AI Companion Products by Intended Usage (n=110)}
      \label{fig:exact}
    \end{figure}
    
    \begin{figure}[H]
      \centering
      \includegraphics[width=1\textwidth]{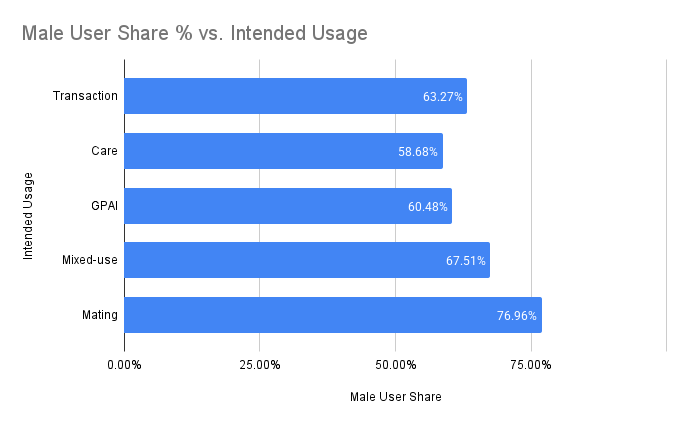}
      \caption{Male User Share vs. Intended Usage}
      \label{fig:exact}
    \end{figure}

    \begin{figure}[H]
      \centering
      \includegraphics[width=1\textwidth]{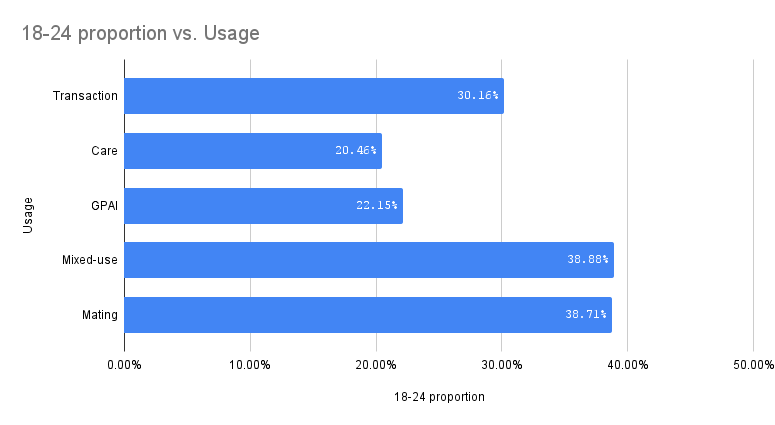}
      \caption{Proportion of 18–24-Year-Old Users Across Age Groups vs. Intended Usage}
      \label{fig:exact}
    \end{figure}

    \begin{figure}[H]
      \centering
      \includegraphics[width=1\textwidth]{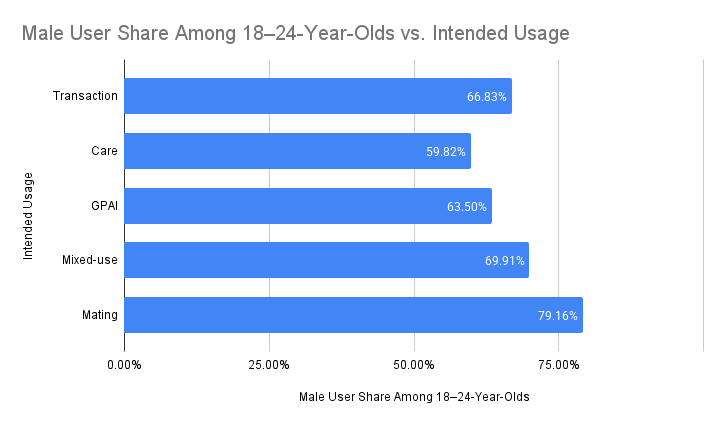}
      \caption{Male User Share Among 18–24-Year-Olds vs. Intended Usage}
      \label{fig:exact}
    \end{figure}

\subsection*{Expanded limitations}
While the market scan provides valuable insights into the growing scale of AI companion markets and potential indications of wide-scale addictions, there are several limitations in our methodology. These limitations are linked to 1) the access to data, 2) the absence of long-term data and 3) the generalisability of metrics. 

\begin{enumerate}
    \item \textbf{Access to Data:} Firstly, our market scan was restricted by access to data. While we attempted to capture a wide range of AI companions through an internet search, the absence of some parasocial AI applications on SimilarWeb and similar platforms meant that we had to exclude some companions. This was especially the case for mobile app-based services, which required a premium SimilarWeb subscription beyond our available resources. Although publicly available data for total app downloads was used, this was only published for apps that reached certain milestones (100K, 1M etc.). AndroidRank was used to provide estimates of total app downloads based on these milestones and the prior trajectory of app downloads over time. For apps not found on this platform but appearing in the top results on the Google Play Store, the latest milestone achieved was given as the number for total downloads. We were also challenged by the lack of financial data for many companies. Many developers of AI companions are likely to be small private companies or startups, which have not disclosed even the most basic financial information. To overcome this challenge, we adopted data imputation to take into account likely undisclosed revenue. We took the minimum disclosed revenue of \$145,000 and applied it across companies with no revenue provided. This technique is limited in that real revenues could be higher or lower, however the main calculations that include revenue are dominated by GPAI and so using imputation compared to not had a less than 10\% effect on these.
    \item \textbf{Absence of long-term data:} Secondly, the absence of long-term data has impeded us from drawing empirical conclusions about the implications of the growing AI companion market and the speed of growth. For example, longitudinal surveys or feedback data on user retention, satisfaction and feedback on each of the AI companions could help us understand the relationships users build with AI companions. Moreover, data on users’ engagement with AI companions, including the frequency, duration and the nature of conversations over time, would be useful in analysing types of AI companions that pose mental health or manipulation risks. 
     \item \textbf{Generalisability of metrics:} Lastly, the generalisability of some metrics we implemented could be limited. OpenAI and Anthropic's 2025 studies were used in our market scan to estimate the proportion of cases GPAIs are used as AI companions. However, the OpenAI study is based on a month-long randomised control trial, where 327 participants were asked to use ChatGPT for at least 5 minutes a day. The study, therefore, may not reflect the natural user behaviour of GPAI. On the other hand, Anthropic's study was based on the analysis of 4.5 million randomly sampled conversations between April 6th to 19th alone. This means that our assumption (i.e. any GPAI platform user will follow the same usage pattern as OpenAI and Anthropic's 2025 studies users) could be debated. 
\end{enumerate}

\end{document}